\documentclass[letter]{sig-alternate}

\usepackage{booktabs}
\usepackage{mathrsfs}
\usepackage{textcomp, graphicx, subfigure}
\usepackage{epsfig}
\usepackage{amssymb}
\usepackage{amsmath}
\usepackage{amsfonts}
\usepackage{algorithmic}
\usepackage{algorithm}
\usepackage{multirow}
\usepackage{cite}
\usepackage{mathtools,bbm}

\begin{document}
% \CopyrightYear{2018}
%  \setcopyright{acmcopyright}
% \conferenceinfo{DAC '18,}{June 24--29, 2018, San Francisco, CA, USA}
% \isbn{978-1-4503-5700-5/18/06}\acmPrice{$15.00}
% \doi{https://doi.org/10.1145/3195970.3196114}

\conferenceinfo{DAC'18,} {June 24--29, 2018, San Francisco, CA, USA}\CopyrightYear{2018} \crdata{978-1-4503-5700-5/18/06\ ...\$15.00 \\ https://doi.org/10.1145/3195970.3196114}

\title{Similarity-Aware Spectral Sparsification by Edge Filtering}

\author{
\alignauthor
       Zhuo Feng\\
      \affaddr{Department of Electrical and Computer Engineering}\\
      \affaddr{Michigan Technological University}\\
      \affaddr{Houghton, MI, 49931}\\
      \email{zhuofeng@mtu.edu}
}

\maketitle

% \vspace{-0.3cm}
\begin{abstract}
In recent years, spectral graph sparsification techniques that can compute ultra-sparse graph proxies have been extensively studied for accelerating various numerical and graph-related applications. Prior nearly-linear-time spectral sparsification methods first extract low-stretch spanning tree from the original graph to form the backbone of the sparsifier, and then recover small portions of spectrally-critical off-tree edges to the spanning tree to significantly improve the approximation quality. However, it is not clear how many off-tree edges should be recovered for achieving a desired spectral similarity level within the sparsifier. Motivated by recent graph signal processing techniques, this paper proposes a similarity-aware spectral graph sparsification framework that leverages  efficient spectral off-tree edge embedding and filtering schemes to construct spectral sparsifiers with guaranteed spectral similarity (relative condition number) level. An iterative graph densification     scheme is introduced to facilitate efficient and effective filtering of off-tree edges  for highly ill-conditioned problems. The proposed method has been validated using various kinds of graphs obtained from public domain sparse matrix collections relevant to VLSI CAD, finite element analysis, as well as social and data networks frequently studied in many machine learning and data mining applications.
\end{abstract}

 \category{B.7.2}{Design Aids}{simulation}[Integrated Circuits]
 \vspace{-0.23cm}
 \terms{Performance, Algorithms, Verification}
  \vspace{-0.23cm}
\keywords{Spectral graph theory, graph partitioning, iterative methods}
\maketitle

% \keywords{Spectral graph theory, graph partitioning, sparse matrix solver}
%  \vspace{-0.3cm}

\section{Introduction}
 Spectral methods are  playing increasingly important roles in  many  graph and numerical applications \cite{teng2016scalable}, such as scientific computing \cite{spielman2014sdd}, numerical optimization \cite{christiano2011flow}, data mining \cite{peng2015partitioning}, graph analytics \cite{koren2003spectral},  machine learning \cite{defferrard2016convolutional}, graph signal processing \cite{shuman2013emerging}, and VLSI computer-aided design \cite{zhiqiang:dac17,zhuo:dac16}. For example, classical spectral graph partitioning (data clustering) algorithms embed original graphs into low-dimensional space using the first few nontrivial eigenvectors of  graph Laplacians and subsequently perform graph partitioning (data clustering) on the low-dimensional graphs to obtain  high-quality solution \cite{peng2015partitioning}.

% To further push the limit of spectral methods for  large graphs, mathematics and theoretical computer science researchers have extensively studied many theoretically-sound research problems related to spectral graph theory.
Recent \emph{{spectral graph sparsification }} research  \cite{spielman2011graph,   batson2012twice, spielman2011spectral, peng2015partitioning,cohen2017almost,Lee:2017} allows computing nearly-linear-sized subgraphs (sparsifiers) that can robustly preserve the spectrum (i.e., eigenvalues and eigenvectors) of the original graph's Laplacian, which immediately leads to a series of theoretically  {{nearly-linear-time}}    numerical and graph algorithms for solving sparse  matrices, graph-based semi-supervised learning (SSL), spectral graph partitioning (data clustering), and max-flow problems \cite{miller:2010focs,  spielman2011spectral, christiano2011flow, spielman2014sdd}.  For example,  sparsified circuit networks allow for developing more scalable computer-aided (CAD) design algorithms for designing large VLSI systems \cite{zhuo:dac16,zhiqiang:dac17}; sparsified social (data) networks enable to more efficiently understand and analyze large social (data) networks \cite{teng2016scalable}; sparsified matrices can be immediately leveraged to  accelerate the solution computation  of large linear system of equations \cite{zhiqiang:iccad17}. To this end, a spectral sparsification algorithm leveraging an edge sampling scheme that sets sampling probabilities proportional to edge effective resistances (of the original graph) has been proposed in \cite{spielman2011graph}.

A practically-efficient, nearly-linear complexity spectral graph sparsification algorithm has been recently introduced in \cite{zhuo:dac16}, which first extracts a ``spectrally critical" spanning tree subgraph as a backbone of the sparsifier, and subsequently recovers a small portion of dissimilar ``spectrally critical" off-tree edges to the spanning tree. However, in many scientific computing and graph-related applications, it is important to compute  spectral graph sparsifiers of desired spectral similarity level:  introducing too few edges may lead to poor approximation of the original graph, whereas too many edges can result in high computational complexity. For example, when using a preconditioned conjugate gradient (PCG) solver  to solve a symmetric diagonally dominant (SDD) matrix for multiple right-hand-side (RHS) vectors, it is hoped the PCG solver would converge to a good solution as quickly as possible,   which usually requires the sparsifier (preconditioner) to be highly spectrally-similar  to the original problem; on the other hand, in many graph partitioning tasks, only the Fielder vector (the first nontrivial eigenvector) of graph Laplacian is needed \cite{spielmat1996spectral}, so even a  sparsifier with much lower spectral similarity will suffice.

This paper introduces a similarity-aware spectral graph sparsification framework that leverages efficient  spectral off-tree edge embedding and filtering schemes  to construct spectral sparsifiers with guaranteed spectral similarity. The   contribution of this work has been summarized as follows:
\begin{enumerate}
  \item We present a similarity-aware spectral graph sparsification framework by leveraging      spectral off-tree edge embedding and filtering schemes that have been motivated by recent graph signal processing techniques \cite{shuman2013emerging}.
    \item An iterative graph densification procedure is proposed to incrementally improve the approximation  of the sparsifier, which enables to flexibly trade off the complexity and spectral similarity of the sparsified graph.
    \item Extensive experiments have been conducted to validate   the proposed method in various numerical and graph-related applications, such as solving sparse SDD matrices, and spectral graph partitioning, as well as simplification of large social and data  networks.
\end{enumerate}

% The rest of this paper is organized as follows.  Section
% \ref{background_sec} provides a brief introduction to   spectral graph sparsification problems. In Section \ref{main_sec}, a similarity-aware spectral graph sparsification framework is described in details.  Section
% \ref{result_sec}  demonstrates extensive experiment results for a variety of real-world, large-scale sparse Laplacian matrix and  graph problems, which is followed by the conclusion of this work in Section \ref{conclusion}.

\section{Spectral Graph Sparsification}\label{background_sec}
Consider a graph $G=(V,E,w)$ with $V$ denoting the vertex (data point) set of the  graph,   $E$ denoting the edge  set of the  graph, and $w$ denoting a weight (similarity) function that assigns positive weights to all edges. The  graph Laplacian $\mathbf{L_G}$  of $G$ is an SDD  matrix defined as follows:
\begin{equation}\label{formula_laplacian}
\mathbf{L_G}(p,q)=\begin{cases}
-w(p,q) & \text{ if } (p,q)\in E \\
\sum\limits_{(p,t)\in E} w(p,t) & \text{ if } (p=q) \\
0 & \text{otherwise }.
\end{cases}
\end{equation}

Spectral graph sparsification \cite{spielman2011spectral} aims to preserve the original graph spectrum within  ultra-sparse subgraphs (graph sparsifiers), which allows preserving not only cuts in the graph but also eigenvalues and eigenvectors of the original graph Laplacian,  distances (e.g. effective resistances) between vertices, low-dimensional graph embedding, etc. Two graphs $G$ and  $P$ are said to be \textbf{\emph{ $\sigma-$spectrally similar}} if for all real vectors $\mathbf{x} \in \mathbb{R}^V$ their quadratic forms satisfy:
\begin{equation}\label{formula_spectral_similar}
\frac{\mathbf{x}^T\mathbf{L_P}\mathbf{x}}{\sigma}\le \mathbf{x}^T\mathbf{L_G}\mathbf{x} \le \sigma \mathbf{x}^T\mathbf{L_P}\mathbf{x}.
\end{equation}
Define the relative condition number to be $\kappa(\mathbf{L_G},\mathbf{L_P})=\lambda_{max}/\lambda_{min}$,
where  $\lambda_{max}$ and $\lambda_{min}$ denote the largest and smallest generalized eigenvalues satisfying:
\begin{equation}\label{formula_eig_perturb0}
\mathbf{L_G}\mathbf{u}=\lambda \mathbf{L_P}\mathbf{u},
\end{equation}
with $\mathbf{u}$ denoting the generalized eigenvector of ${\lambda}$. It can be further shown that $\kappa(\mathbf{L_G},\mathbf{L_P})\le\sigma^2$, which indicates that a smaller relative condition number or $\sigma^2$ corresponds to a higher spectral similarity. Obviously, we can simply use $\sigma^2$   to denote the upper bound of the relative condition number.

\section{Similarity-Aware Spectral Sparsification By Edge Filtering}\label{main_sec}
\subsection{Overview of Our Approach}

\begin{figure}
\centering \epsfig{file=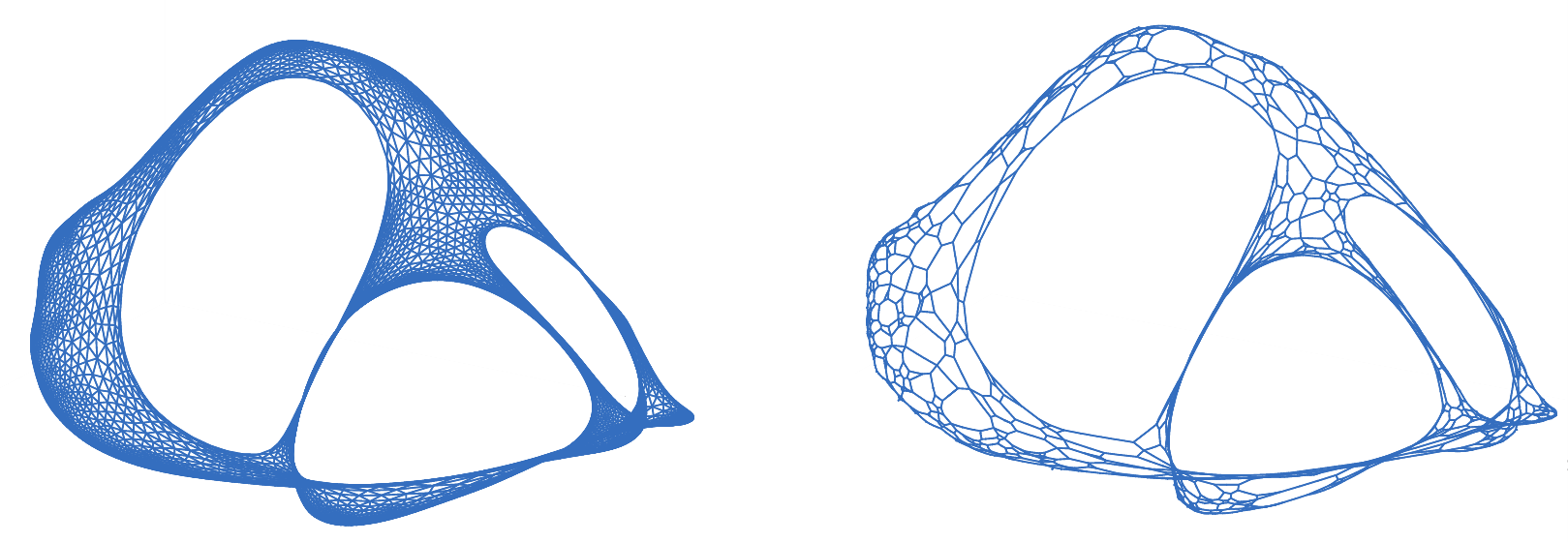, scale=0.5} \caption{Two  spectrally-similar airfoil graphs. \protect\label{fig:similarity}}
\end{figure}

% \begin{figure}
% \centering \epsfig{file=flowchart.eps, scale=0.35} \caption{Similarity-aware spectral sparsification of graphs. \protect\label{fig:flowchart}}
% \end{figure}
The overview of the proposed method for similarity-aware spectral sparsification of undirected graphs has been summarized as follows. For a given input graph, the following key procedures are involved in the proposed algorithm flow: \textbf{(a)}  low-stretch spanning tree \cite{elkin2008lst,abraham2012}  extraction based on its original  graph Laplacian; \textbf{(b)} spectral (generalized eigenvalue)  embedding and filtering of off-tree edges by leveraging the  recent spectral  perturbation analysis framework \cite{zhuo:dac16}; \textbf{(c)} incremental sparsifier improvement (graph densification) by gradually adding small portions of dissimilar off-tree edges  to the spanning tree. Fig.~\ref{fig:similarity} shows the spectral drawings \cite{koren2003spectral} of an airfoil graph \cite{davis2011matrix} as well as its spectrally-similar subgraph computed by the proposed similarity-aware spectral sparsification algorithm.

%   Since recovering too many off-tree edges will result in high computational costs, whereas recovering insufficient amount of  edges can lead to poor or  misleading approximation results,  in this work we propose an incremental graph densification procedure leveraging an efficient off-edge filtering scheme.

In the rest of this paper,  we assume that $G=(V,E,w)$ is a weighted, undirected and connected graph, whereas $P=(V,E_s,w_s)$ is its sparsifier. To simplify the our analysis, we assume the edge weights in the sparsifier remain the same as the original ones, though  edge   re-scaling schemes \cite{spielman2011spectral}  can be applied to further improve the approximation. The descending   eigenvalues of $\mathbf{L^+_P L_G}$ are denoted by $\lambda _{max}={\lambda _1} \ge {\lambda _2} \ge  \cdots  \ge {\lambda _n} \ge 1$, where $\mathbf{L^+_P}$ denotes the  Moore-Penrose  pseudoinverse of $\mathbf{L_P}$.
\subsection{Spectral Embedding of Off-Tree Edges }
% \subsubsection{Spectral Distortion of The Subgraph.}
It has been shown that there are  not too many large generalized eigenvalues for spanning-tree sparsifiers \cite{spielman2009note}: $\mathbf{L^+_P} \mathbf{L_G}$  has at most $k$ generalized eigenvalues greater than ${\textstyle{{{\rm{s}}{{\rm{t}}_{P}}\left( G \right)} \over k}}$, where   $\textstyle{{\rm{s}}{{\rm{t}}_{P}}}\left( G \right)$ denotes the total stretch of the spanning-tree subgraph $P$ with respect the original graph $G$. Recent  research results show that every undirected   graph has a low-stretch spanning tree (LSST) such that \cite{elkin2008lst,abraham2012}:
\begin{equation}\label{formula_stretch}
O(m \log n \log \log n)\geq{{\rm{s}}{{\rm{t}}_{P}}\left( G \right)} = \mathbf{Trace(L^+_P L_G})= \sum\limits_{i = 1}^{n} {{\lambda _i}},
\end{equation}
where $m=|E|$ and $n=|V|$. As a result, it is possible to construct an ultra-sparse yet spectrally similar sparsifier  by  recovering only a small portion of  important off-tree edges to the spanning tree: for example,    $\sigma$-similar spectral sparsifiers with   $O(\frac{m \log n \log \log n}{\sigma^2})$ off-tree edges can be computed efficiently using perturbation-based method \cite{zhuo:dac16}.

% \subsubsection{Identification of Key Off-Tree Edges.}
To identify  important off-tree edges the following generalized eigenvalue perturbation analysis is considered \cite{zhuo:dac16}:
\begin{equation}\label{formula_eig_perturb1}
\mathbf{L_G}\left( {\mathbf{u_i} + \delta \mathbf{u_i}} \right) = \left( {{\lambda _i} + \delta {\lambda _i}} \right)\left( \mathbf{L_P + \delta L_P} \right)\left( {\mathbf{u_i} + \delta \mathbf{u_i}} \right),
\end{equation}
where a perturbation matrix $\delta \mathbf{L_P}$ is applied for the inclusion of extra off-tree edges into $\mathbf{L_P}$ and results in perturbed generalized eigenvalues and eigenvectors  ${\lambda _i} + \delta {\lambda _i}$ and $\mathbf{u_i} + \delta \mathbf{u_i}$ for $i=1,...,n$, respectively. The key to effective spectral sparsification is  to identify the key off-tree edges that will result in the greatest reduction in dominant generalized eigenvalues.  To this end,  the following scheme for embedding generalized eigenvalues into each  off-tree edge  is adopted in this work \cite{zhuo:dac16}: \\
 \noindent\textbf{Step 1:} Start with  an initial random vector $\mathbf{h_0} = \sum\limits_{i = 1}^{n} {{\alpha _i}\mathbf{u_i}}$, where  $\mathbf{u_i}$ are the $\mathbf{L_P}$-orthogonal generalized eigenvectors of $\mathbf{L^+_P L_G}$  that satisfy $\mathbf{u_i^T L_Pu_j}^{} = 1$ for $i=j$, and $0$ for $i\ne j$; \\
 \noindent\textbf{Step 2:} Perform  $t$-step generalized power iterations with $\mathbf{h_0}$ to obtain $\mathbf{h_t} = \left(\mathbf{L_P^{+}L_G}\right)^t\mathbf{h_0} = \sum\limits_{i = 1}^{n} {{\alpha _i}{\lambda^t _i}\mathbf{u_i}}$; $\mathbf{h_t}$ will be a good approximation of dominant eigenvectors; \\
 \noindent\textbf{Step 3:} Compute the Laplacian quadratic form  for $\delta \mathbf{L_{P, {max}}}= \mathbf{L_G-L_P}$ with $\mathbf{h_t}$:
  \begin{equation}\label{formula_pwr_iter5new}
  \begin{array}{l}
Q\mathbf{_{\delta L_{P, {max}}}(h_t)={h_t}^T\delta L_{P, {max}}{h_t}} = \sum\limits_{i = 1}^{n} {{{\left( {{\alpha _i}{\lambda^t _i}} \right)}^2(\lambda_i-1)}}\\=\sum\limits_{(p,q)\in E\setminus E_s}^{} w_{p,q}\sum\limits_{i = 1}^{n}{\alpha^2_i}{\lambda^{2t} _i}\left(\mathbf{u_i}^T \mathbf{e}_{p,q}\right)^2=\sum\limits_{(p,q)\in E\setminus E_s}^{} heat_{(p,q)},
\end{array}
\end{equation}
where  $\mathbf{\delta L_{P, {max}}}$ denotes the perturbation of  $\mathbf{ L_{P}}$  including all off-tree edges,
$\mathbf{e}_{p,q}\in \mathbb{R}^n$ denotes the vector that has   the $p$-th element being $1$, the $q$-th element being $-1$ and others being $0$, and $heat_{(p,q)}$ denotes the edge Joule heat of the off-tree edge $(p,q)$. The amplitude of $Q\mathbf{_{\delta L_{P, {max}}}(h_t)}$ reflects the spectral similarity between graphs $P$ and $G$: larger $Q\mathbf{_{\delta L_{P, {max}}}(h_t)}$  indicates greater $\sigma^2$ and thus lower spectral similarity.  More importantly, (\ref{formula_pwr_iter5new}) allows  embedding generalized eigenvalues into the Laplacian quadratic form of each off-tree edge and  ranking  each off-tree edge according to  its edge Joule heat (spectral criticality): recovering the off-tree edges  with largest  $heat_{(p,q)}$     will most significantly decrease the  largest generalized eigenvalues. In practice, using a small number (e.g. $t=2$) of generalized power iterations will suffice for spectral edge embedding purpose.

\subsection{``Spectrally-Unique'' Off-Tree Edges}
 To simplify the following analysis, we define a ``spectrally unique" off-tree edge $e_i$ to be the one that connects to vertices $p_i$ and $q_i$, and only impacts a single  large generalized eigenvalue $\lambda_i$. Then the  truncated version of (\ref{formula_pwr_iter5new}) including the top $k$  dominant ``spectrally-unique"  off-tree edges for fixing the top $k$ largest eigenvalues of $\mathbf{L^+_P L_G}$ can be expressed as follows for $\lambda_i \gg 1$:
 \begin{equation}\label{formula_eigfix}
 \begin{array}{l}
Q\mathbf{_{\delta L_{P, {max}}}(h_t)} \approx \sum\limits_{i = 1}^{k} w_{p_i,q_i}{\alpha^2_i}{\lambda^{2t}_i}\left(\mathbf{u_i}^T  \mathbf{e}_{p_i,q_i }\right)^2\approx\sum\limits_{i = 1}^k {{{ {{\alpha^2 _i}{\lambda^{2t+1} _i}}}}}.
\end{array}
\end{equation}
Since each off-tree edge   only impacts  one generalized eigenvalue, we can express $\mathbf{e}_{p_i,q_i}=\gamma_i \mathbf{L_P u_i}$ according to (\ref{formula_pwr_iter5new}), which leads to:
 \begin{equation}\label{formula_p-orth3}
\mathbf{u_j^T e}_{p_i,q_i}= \left\{ \begin{array}{l}
\gamma_i,i = j,\\
0,i \ne j.
\end{array} \right.\end{equation}
%which also indicates that:
%\begin{equation}\label{formula_p-orth2}
%e_{p_i,q_i}=\gamma_i L_P u_i.
%\end{equation}
Then the effective resistance of edge $e_i$ in $P$ becomes:
  \begin{equation}\label{formula_Reff}
R^{\mathbf{eff}}_{e_i}=\mathbf{e}^T_{p_i,q_i}\mathbf{L^+_P e}_{p_i,q_i}=\gamma^2_i \mathbf{u^T_i L_P u_i}=\gamma^2_i,
\end{equation}
%then the stretch of edge $e_i$  according to (\ref{formula_stretch_edge}) becomes:
%  \begin{equation}\label{formula_stretch2}
%{{\rm{s}}{{\rm{t}}_{P}}\left( e_i \right)}= w_{p_i,q_i} \gamma^2_i,
%\end{equation}
which immediately leads to:
  \begin{equation}\label{formula_eigfix2}
Q\mathbf{_{\delta L_{P, {max}}}(h_t)} \approx  \sum\limits_{i = 1}^{k}  {\alpha^2_i}{\lambda^{2t}_i} w_{p_i,q_i} R^{\mathbf{eff}}_{e_i}\approx\sum\limits_{i = 1}^k {{{ {{\alpha^2 _i}{\lambda^{2t+1} _i}} }}}.
\end{equation}
Since the stretch of off-tree edge $e_i$ is computed by ${{\rm{s}}{{\rm{t}}_{P}}\left( e_i \right)}=w_{p_i,q_i} R^{\mathbf{eff}}_{e_i}$, (\ref{formula_eigfix2}) also indicates that ${{\rm{s}}{{\rm{t}}_{P}}\left( e_i \right)}\approx\lambda_i$ holds for ``spectrally-unique" off-tree edges. Consequently, the key off-tree edges identified by (\ref{formula_pwr_iter5new}) or (\ref{formula_eigfix2}) will have the largest stretch values  and therefore most significantly impact the largest eigenvalues of $\mathbf{L_P^{+}L_G}$.  (\ref{formula_eigfix2}) also can be considered as a randomized version of  $\mathbf{Trace({L_P^{+}}L_G)}$ that is further scaled up by a factor of $\lambda^{2t}_{i}$.

\subsection{Spectral Sparsification as A Graph Filter}
 Although  (\ref{formula_pwr_iter5new}) and  (\ref{formula_eigfix2})  provide  a  spectral ranking for each off-tree edge, it is not clear how many off-tree  edges should be recovered to the spanning tree for achieving a desired spectral similarity level. To this end, we introduce  a simple yet effective spectral off-tree edge filtering scheme motivated by recent graph signal processing techniques \cite{shuman2013emerging}.
To more efficiently analyze signals on general undirected graphs,   graph signal processing techniques have been extensively studied recently \cite{shuman2013emerging}. There is a clear analogy between traditional signal processing based on classical Fourier analysis and graph signal processing: 1) the signals at different time points in classical Fourier analysis correspond to the signals at different  nodes in an undirected graph; 2) the more slowly oscillating functions   in time domain correspond to the graph Laplacian eigenvectors associated with lower eigenvalues and more slowly varying (smoother) components across the graph.
% For example, the first  few nontrivial eigenvectors associated with the smallest non-zero eigenvalues of a path graph Laplacian have been illustrated in Fig. \ref{fig:linegraph}, where the increasing eigenvalues correspond to increasing oscillation frequencies in the line graph.
A comprehensive review of fundamental signal processing operations, such as filtering, translation, modulation, dilation, and down-sampling to the graph setting has been provided in \cite{shuman2013emerging}.

% \begin{figure}
% \centering \epsfig{file=linegraph.eps, scale=0.3} \caption{The  eigenvectors ($v_i$) of increasing eigenvalues ($\tau_i$) for a path graph. \protect\label{fig:linegraph}}
% \end{figure}

Spectral sparsification  aims to  maintain a simplest subgraph  sufficient for preserving the slowly-varying  or ``low-frequency" signals on graphs, which therefore can be regarded as a ``low-pass" graph filter. In other words, such spectrally sparsified graphs will be able to preserve the eigenvectors associated with low eigenvalues more accurately than high eigenvalues, and thus will retain ``low-frequency" graph signals sufficiently well, but  not so well for highly-oscillating (signal) components due to the  missing edges.

In practice, preserving the spectral (structural) properties  of the original graph within the spectral sparsifier is  key to design of  many fast  numerical and graph-related  algorithms \cite{spielman2011graph,miller:2010focs, christiano2011flow, spielman2014sdd}. For example, when using spectral sparsifier as a preconditioner in preconditioned conjugate gradient (PCG) iterations, the convergence rate only depends on the  spectral similarity (or relative condition number) for achieving a desired accuracy level, while in spectral graph partitioning and data clustering tasks only the first few eigenvectors associated with the smallest nontrivial eigenvalues of graph Laplacian are needed \cite{spielmat1996spectral,peng2015partitioning}.

 \subsection{Off-Tree Edge Filtering with Joule Heat}
  To only recover the off-tree edges that are most critical for achieving the desired spectral similarity level, we propose the following scheme for truncating ``spectrally-unique" off-tree edges based on each edge's Joule heat. For a spanning-tree preconditioner, since there will be at most $k$ generalized eigenvalues that are greater than ${{\rm{s}}{{\rm{t}}_{P}}\left( G \right)} /k$, the following simple yet nearly worst-case generalized eigenvalue distribution can be assumed:
%    \begin{equation}\label{formula_eig_dist}
% \lambda _{i}= \frac{2^{\sgn(i-1)}\lambda_{max}}{i+{\sgn(i-1)}}=\left\{ \begin{array}{l}
% \frac{{\rm{s}}{{\rm{t}}_{P}}\left( G \right)}{2}=\lambda_{max},i = 1,\\
% \frac{{\rm{s}}{{\rm{t}}_{P}}\left( G \right)}{i+1}=\frac{2\lambda_{max}}{i+1},i \geq 2.
% \end{array} \right.\end{equation}

\begin{equation}\label{formula_eig_dist}
\lambda _{i}= \frac{2\lambda_{max}}{i+1}=\frac{{\rm{s}}{{\rm{t}}_{P}}\left( G \right)}{i+1},i\geq 1.
\end{equation}
To  most economically select  the top-$k$ ``spectrally-unique"  off-tree edges that will dominantly impact the top-$k$ largest generalized eigenvalues, the  following sum of quadratic forms  (Joule heat levels) can be computed based on (\ref{formula_eigfix2}) by performing $t$-step generalized power iterations with $r$ multiple random vectors  $\mathbf{h_{t,{1}},...,h_{t,{r}}}$:
\begin{equation}\label{formula_pwr_iter6}
Q\mathbf{_{\delta L_{P, {max}}}(h_{t,{1}},...,h_{t,{r}})}   \approx  \sum\limits_{j = 1}^r \sum\limits_{i = 1}^k {{{(\alpha_{i,j})^2\left( \frac{2\lambda_{max}}{i+1} \right)}^{2t+1}}}.
\end{equation}
The goal is to select top $k$  ``spectrally-unique"  off-tree edges  for fixing the top  $k$ largest generalized eigenvalues such that  the resulting upper bound of the relative condition number will become ${\sigma}^2=\frac{\tilde{\lambda}_{max}}{\tilde{\lambda}_{min}}$, where $\tilde{\lambda}_{max}$ and $\tilde{\lambda}_{min}$ denote the largest and smallest eigenvalues of $\mathbf{L_P^+L_G}$ after adding top-k ``spectrally-unique" off-tree edges. Then we have:
\begin{equation}\label{formula_eig_k}
k=2\lambda_{max}/\tilde{\lambda}_{max}-1.
\end{equation}
When using multiple random vectors for computing (\ref{formula_pwr_iter6}), it is expected that ${{\sum\limits_{j = 1}^r \alpha _{k,j}^2}}\approx {{\sum\limits_{j = 1}^r \alpha _{1,j}^2}}$, which allows us to define the normalized edge Joule heat $\theta _k$ for the $k$-th ``spectrally-unique"  off-tree edge  through the following simplifications:
  \begin{equation}\label{formula_heatratio}
 {\theta _k} = \frac{{{heat_{{\lambda _k}}}}}{{{heat_{{\lambda _1}}}}} = {\left( {\frac{{{\sum\limits_{j = 1}^r \alpha _{k,j}^2}}}{{{\sum\limits_{j = 1}^r \alpha^2 _{1,j}}}}} \right)}{\left( \frac{\tilde{\lambda}_{max} }{ \lambda_{max}} \right)^{2t+1}}\approx\left( \frac{{\sigma}^2  \tilde{\lambda}_{min} }{ \lambda_{max}} \right)^{2t+1}.
\end{equation}
The key idea of the proposed similarity-aware spectral sparsification is to leverage the normalized Joule heat (\ref{formula_heatratio}) as a  threshold for filtering off-tree edges: only the off-tree edges with normalized Joule heat values greater than $\theta _k$ will be selected for inclusion into the spanning tree for achieving the desired spectral similarity  ($\sigma$) level.
Although the above  scheme is derived for filtering ``spectrally-unique" off-tree edges,   general off-tree edges also can be filtered using similar strategies. Since adding the off-tree edges with largest Joule heat to the subgraph will mainly impact the largest generalized eigenvalues but not the smallest ones, we will  assume $\tilde{\lambda}_{min}\approx {\lambda}_{min}$, and  use the following edge truncation scheme for filtering general off-tree edges: the off-tree edge $(p,q)$ will be included into the sparsifier if its normalized Joule heat value is greater than the  threshold determined by:
\begin{equation}\label{formula_heatratio_final}
\theta _{(p,q)}=\frac{heat_{(p,q)} }{heat_{max}} \ge \theta_{{\sigma}} \approx \left( \frac{{\sigma}^2 {\lambda}_{min}}{ \lambda_{max}} \right)^{2t+1},
\end{equation}
where $\theta_{{\sigma}}$ denotes the threshold for achieving the $\sigma-$spectral similarity in the sparsifier, and $heat_{max}$ denotes the maximum  Joule heat  of all off-tree edges computed by (\ref{formula_pwr_iter5new}) with multiple initial random vectors.

\subsection{Estimation of Extreme Eigenvalues}\label{sec_extreme-eigvalue}
To achieve the above spectral off-tree edge filtering scheme, we need to compute $\theta_{{\sigma}} $ in (\ref{formula_heatratio_final}) that further requires to estimate the extreme    eigenvalues  $\lambda_{max}$ and $\lambda_{min}$ of $\mathbf{L_P^+L_G}$. In this work, we propose the following efficient methods for computing these extreme generalized eigenvalues.
\subsubsection{Estimating  $\lambda_{max}$ via Power Iterations }
 Since generalized power iterations  converge at a geometric rate determined by the separation of the two largest generalized eigenvalues $\lambda_{max}=\lambda_1>\lambda_2$ , the error of the estimated  eigenvalue  will decrease quickly when $|\lambda_2/\lambda_1|$ is small. It has been shown that the largest eigenvalues of $\mathbf{L_P^{+}L_G}$ are well separated from each other \cite{spielman2009note}, which thus leads to very fast convergence of generalized power iterations for estimating  $\lambda_{1}$. To achieve scalable performance of power iterations, we can adopt  recent graph-theoretic algebraic multigrid (AMG) methods for  solving the sparsified Laplacian matrix $\mathbf{L_P}$  \cite{livne2012lean, zhiqiang:iccad17}.

\subsubsection{Estimating  $\lambda_{min}$ via Node Coloring}
Since the smallest eigenvalues of $\mathbf{L_P^{+}L_G}$ are crowded together \cite{spielman2009note}, using (shifted) inverse power iterations may not be efficient due to the extremely slow convergence rate. To the extent of our knowledge, none of existing eigenvalue decomposition methods can  efficiently compute $\lambda_{min}$.

This work exploits the following Courant-Fischer theorem for generalized eigenvalue problems:
\begin{equation}\label{formula_courant-fischer}
\lambda_{min}=\min_{|\mathbf{x}| \neq 0}\frac{\mathbf{x^T L_G {x}}}{\mathbf{x^T L_P {x}}},
\end{equation}
where $\mathbf{x}$ is also required to be orthogonal to the all-one vector. (\ref{formula_courant-fischer}) indicates that if we can find a vector $\mathbf{x}$ that minimizes the ratio between the quadratic forms of the original and sparsified Laplacians, $\lambda_{min}$ can be subsequently computed. By restricting the values in $\mathbf{x}$ to be only  $1$ or $0$, which can be considered as assigning one of the two colors to each node in graphs $G$ and $P$,  the following   simplifications can be made:
\begin{equation}\label{formula_courant-fischer-2}
\lambda_{min}\approx\mathop{\min_{\mathbf{|x|} \neq 0}}_{x(i)\in \left\{0,1 \right\}}\mathbf{\frac{x^T L_G x}{x^T L_P x}}=\mathop{\min_{\mathbf{|x|} \neq 0}}_{x(i)\in \left\{0,1 \right\}}{\frac{\sum\limits_{x(p)\neq x(q), (p,q)\in E} w_{pq}}{\sum\limits_{x(p)\neq x(q), (p,q)\in E_s} w_{pq}}},
\end{equation}
which will always allow estimating an upper bound for $\lambda_{min}$.  To this end, we first initialize all nodes with $0$ value and subsequently try to find a node $p$ such that  the ratio between quadratic forms can be minimized:
\begin{equation}\label{formula_courant-fischer-3}
\lambda_{min}\approx \min_{p\in V}{\frac{ L_G(p,p)}{L_P(p,p)}}.
\end{equation}
The above procedure for estimating $\lambda_{min}$ only requires finding the node with the smallest node degree ratio and thus can be easily implemented and efficiently performed for even very graph problems.  Our  results for real-world  graphs show that the proposed method is highly efficient and can reasonably estimate the smallest generalized eigenvalues when compared with existing generalized eigenvalue  methods \cite{saad2011eigbook}.
\subsection{Iterative Graph Densification}\label{sec_iter_sparse}
To achieve more effective edge filtering for  similarity-aware spectral graph sparsification, we propose to iteratively recover off-tree edges to  the sparsifier through an incremental graph densification procedure. Each densification iteration adds a small portion of ``filtered" off-tree edges to the latest spectral sparsifier, while the  spectral similarity is estimated to determine if more off-tree edges are needed. The  $i$-th graph densification iteration includes the following steps:
\begin{enumerate}
   \item Update the subgraph Laplacian matrix $\mathbf{L_P}$ as well as its    solver by leveraging recent graph-theoretic algebraic multigrid methods \cite{livne2012lean,zhiqiang:iccad17};
   \item Estimate the spectral similarity  by computing $\lambda_{max}$ and $\lambda_{min}$ using the methods described in Section \ref{sec_extreme-eigvalue};
  \item If the spectral similarity is not satisfactory, continue with the following steps; otherwise, terminate the subgraph densification procedure.
 \item Perform $t$-step generalized power iterations with  $O(\log |V|)$ random vectors to compute the sum of Laplacian quadratic forms  (\ref{formula_pwr_iter6});
  \item Rank and filter each off-tree edge  according to its normalized Joule heat value using  the threshold  $\theta_{{\sigma}} $ in (\ref{formula_heatratio_final});
  \item  Check the similarity of each selected off-tree edge and only add dissimilar edges to the latest sparsifier.
\end{enumerate}

 \vspace{-0.3cm}

\section{Experimental results}\label{result_sec}
The proposed  spectral  sparsification algorithm has been implemented in $C$++ \footnote{https://sites.google.com/mtu.edu/zhuofeng-graphspar}.
The test cases used in this paper have been selected from a great variety of  matrices that have been used in circuit simulation, finite element analysis, machine learning and data mining applications. If the original matrix is not a graph Laplacian, it will be converted into a graph Laplacian by setting each edge weight using the absolute value of each nonzero entry in the lower triangular matrix; if edge weights are not available in the original matrix file, a unit edge weight will be assigned to all edges. All of our experiments have been conducted using a single CPU core of a computing platform running 64-bit RHEW 7.2 with a $2.67$GHz 12-core CPU and $50$ GB memory.

\subsection{Estimation of Extreme Eigenvalues}
\begin{table}
\begin{center}
 \addtolength{\tabcolsep}{-2.5pt} \centering
\caption{Results of extreme  eigenvalue estimations.}
\scriptsize\begin{tabular}{|c|c|c|c|c|c|c|}
 \hline   Test Cases& $\lambda_{min}$ & $\tilde\lambda_{min}$ &$\epsilon_{\lambda_{min}}$  &  $\lambda_{max}$& $\tilde\lambda_{max}$&$\epsilon_{\lambda_{max}}$ \\
  \hline  fe\_rotor & $1.34$ & $1.40$ &$4.4\%$  &  $120.9$& $116.7$&$3.5\%$ \\
   \hline  pdb1HYS & $1.71$ & $1.89$ &$10.5\%$  &  $120.6$& $113.2$&$6.1\%$ \\
      \hline  bcsstk36 & $1.18$ & $1.27$ &$7.6\%$  &  $96.0$& $92.4$&$3.8\%$ \\
            \hline  brack2 & $1.15$ & $1.20$ &$4.3\%$  &  $92.6$& $90.3$&$2.5\%$ \\
 \hline  raefsky3 & $1.13$ & $1.25$ &$10.5\%$  &  $84.4$& $82.7$&$2.0\%$ \\
   \hline
\end{tabular}\label{table:eigvalue}
\end{center}
\end{table}

In Table \ref{table:eigvalue}, the extreme generalized  eigenvalues ($\tilde\lambda_{min}$ and $\tilde\lambda_{max}$) estimated by the proposed methods (Section \ref{sec_extreme-eigvalue}) are compared with the   ones ($\lambda_{min}$ and $\lambda_{max}$)  computed by the ``\emph{eigs}" function in Matlab for sparse matrices in \cite{davis2011matrix}, while the relative errors ($\epsilon_{\lambda_{min}}$ and $\epsilon_{\lambda_{max}}$) are also shown.  $\tilde\lambda_{max}$ is estimated using less than ten generalized power iterations.

\begin{figure}
\centering \epsfig{file=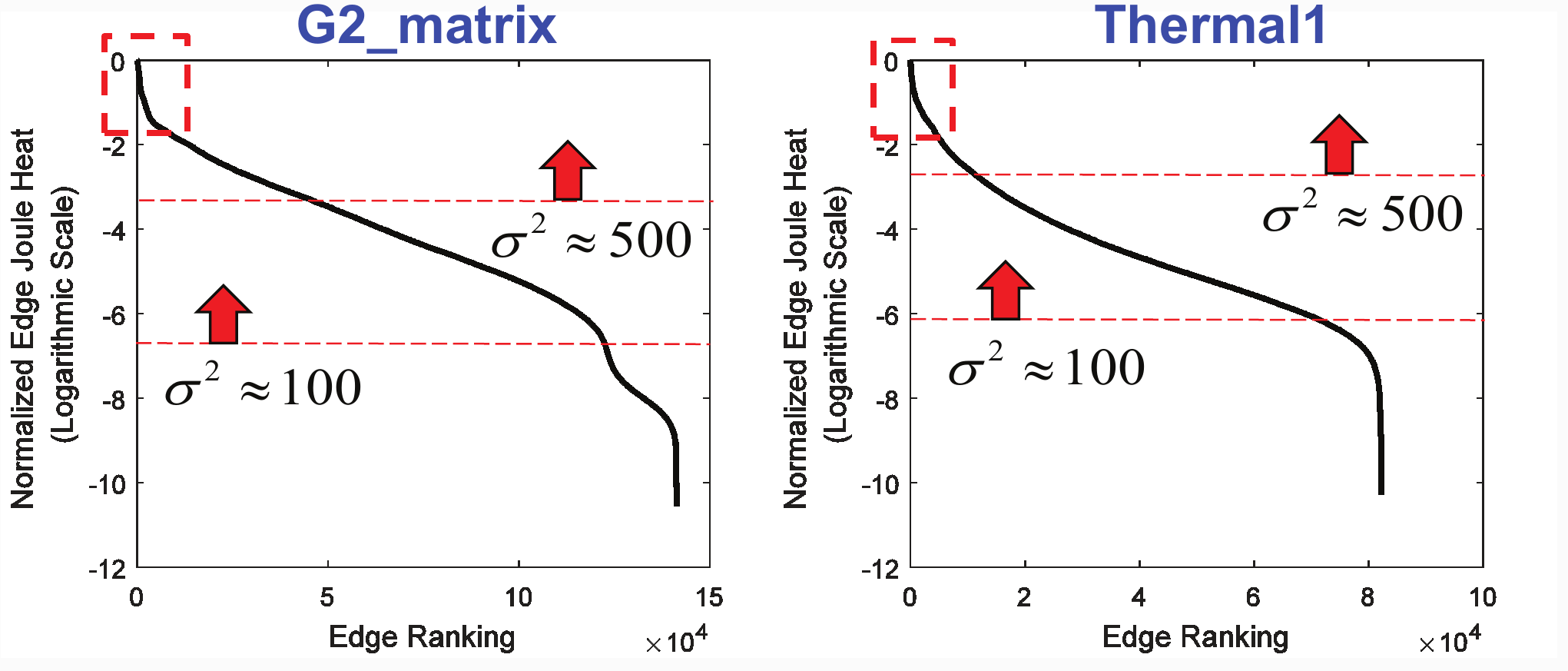, scale=0.43} \caption{Spectral edge  ranking and filtering by normalized Joule heat of off-tree edges for $G2\_circuit$ (left)  and $Thermal1$ (right) test cases\cite{davis2011matrix} with  top off-tree edges highlighted in red rectangles. \protect\label{fig:edge_rank}}
\end{figure}

We also illustrate the results of  spectral edge ranking and filtering according to Joule heat levels computed by  one-step generalized power iteration using (\ref{formula_pwr_iter5new})   in Fig. \ref{fig:edge_rank}  for two  sparse matrices in \cite{davis2011matrix}. The thresholds of normalized edge Joule heat values required for spectral edge filtering are labeled using red dash lines. It is observed in Fig. \ref{fig:edge_rank} there is a sharp change of the top normalized edge Joule heat values, which indicates that there are not  many large eigenvalues of $\mathbf{L_P^{+}L_G}$ in both cases and agrees well with the prior theoretical analysis \cite{spielman2009note}.

 \vspace{-0.13cm}
\subsection{A Scalable Sparse SDD Matrix Solver}
The spectral sparsifier obtained by the proposed similarity-aware algorithm  is also leveraged as a preconditioner in a PCG solver.  The RHS input vector $b$ is generated randomly and the solver is set to converge to  an accuracy level $||A x-b||<10^{-3}||b||$  for all test cases.  ``$|V|$" and ``$|E|$" denote the numbers of nodes and edges in the original graph, whereas ``$|E_{\sigma^2}|$",  ``$N_{\sigma^2}$" and ``$T_{\sigma^2}$" denote the number of edges in the sparsifier, the number of PCG iterations required for converging to the desired accuracy level, and the total time of graph sparsification for achieving the spectral similarity of $\sigma^2$, respectively. As observed in all test cases, there are very clear trade-offs between the graph density,  computation time,   and spectral similarity for all spectral sparsifiers extracted using the proposed method: sparsifiers with higher spectral similarities (smaller $\sigma^2$)  allow converging to the required solution accuracy level in much fewer PCG iterations, but need to retain more edges in the subgraphs and thus require longer time to compute (sparsify).

\begin{table}
\begin{center}
\tiny \addtolength{\tabcolsep}{-2.5pt} \centering
\caption{Results of iterative SDD matrix solver.}
\begin{tabular}{|c|c|c|c|c|c|c|c|c|}
 \hline   Graphs & $|V|$ & $|E|$ & $\frac{|E_{50}|}{|V|}$ & $N_{50}$ & $T_{50}$ &  $\frac{|E_{200}|}{|V|}$ & $N_{200}$ &  $T_{200}$\\
 \hline G3\_circuit & 1.6E6 &  3.0E6 & 1.11 & 21   & 20s & 1.05& 37& 8s \\
 \hline thermal2 & 1.2E6 &  3.7E6 & 1.14  & 20  & 23s & 1.06 & 36 & 9s\\
 \hline ecology2 & 1.0E6 &   2.0E6 & 1.14  & 20  & 16s &  1.06  & 40  & 5s \\
 \hline tmt\_sym & 0.7E6 & 2.2E6 & 1.21  & 19  & 16s &  1.14  & 38  & 4s\\
 \hline paraboli\_fem & 0.5E6 & 1.6E6 &  1.22  & 18  & 16s &  1.09  & 38  & 3s\\
  \hline
\end{tabular}\label{table:uflsdd}
\end{center}
\end{table}
%  Similar runtime  scalability is observed from Table \ref{table:uflsdd} for solving sparse matrices from \cite{davis2011matrix}. In all  test cases, the proposed spectral graph sparsification algorithm can find  tree-like ultra-sparsifiers with high spectral similarity.
%\vspace{-0.101500cm}
%

%  Our   results show that the proposed method can extract the whole spectral sparsifier (spanning tree and ultra-sparsifier) in nearly-linear time, as shown in Fig. \ref{fig:runtimeplot}. It should be noted that each spectral sparsifier needs to be extracted once and can be reused or incrementally updated  many times \cite{xueqian:tcad15,lengfei:tcad15}.
% \begin{figure}
% \centering \epsfig{file=runtimeplot.eps, scale=0.85}\caption{Total runtime of the spectral graph sparsification procedure for sparse SDD matrices from \cite{davis2011matrix}. \protect\label{fig:runtimeplot}}
% \end{figure}
 \vspace{-0.03cm}
\subsection{A Scalable Spectral Graph Partitioner}
It has been shown that by applying only a few inverse power iterations, the approximate Fiedler vector ($u_{f}$) that corresponds to the smallest nonzero eigenvalue  of the (normalized) graph Laplacian matrix can be obtained for obtaining high-quality graph partitioning solution \cite{spielman2014sdd}. Therefore, using the spectral sparsifiers computed by the proposed spectral  sparsification algorithm can immediately accelerate the PCG solver for inverse power iterations, leading to scalable performance for  graph partitioning problems \cite{spielman2014sdd}. In fact, if the spectral  sparsifier is already a good approximation of the original graph, its Fiedler vector can be directly used for   partitioning  the original graph.

We implement the accelerated  spectral graph partitioning algorithm, and test it with sparse matrices in \cite{davis2011matrix} and several 2D mesh graphs synthesized with random edge weights. As shown in Table \ref{table:fiedler}, the graphs associated with sparse matrices have been partitioned into two pieces using sign cut method \cite{spielmat1996spectral}  according to the approximate Fiedler vectors computed by a few steps of inverse power iterations. The direct solver \cite{cholmod}  and the preconditioned iterative solver are invoked within each inverse power iteration for updating the approximate Fiedler vectors $u_{f}$ and $\tilde u_{f}$, respectively.  ``$\frac{|V_+|}{|V_-|}$" denotes the ratio of nodes assigned with positive and negative signs according to the approximate Fiedler vector, and ``Rel.Err." denotes the relative error of the proposed solver compared to the direct solver computed by $\frac{|V_{dif}|}{|V|}$, where $|V_{dif}|$ denotes the number of nodes with different signs in $u_{f}$ and $\tilde u_{f}$. `` $T_{D}$" (`` $T_{I}$") and ``$M_{D}$" (`` $M_{I}$") denote   the total solution time (excluding   sparsification time) and   memory cost of  the direct (iterative) method. We extract sparsifiers with $\sigma^2\leq 200$ for all test cases.

%  It can be observed that the proposed  preconditioned spectral graph partitioner  only results in a very small portion of nodes (0.07\% to 4\%)   assigned with different signs when comparing with the original spectral graph partitioner, while achieving significant runtime and memory  savings ($4-10\times$). The approximate Fiedler vector computed by our fast solver for the test case ``mesh\_1M" is also illustrated in Fig. \ref{fig:fiedlercomp}, showing rather good agreement with the true solution.
% \begin{figure}
% \centering \epsfig{file=fiedlervec_sol.eps, scale=0.485}\epsfig{file=fiedlervec_err.eps, scale=0.485} \caption{The approximate Fiedler vector (left)  and its magnitude error (right) for ``mesh\_1M". \protect\label{fig:fiedlercomp}}
% \end{figure}

\begin{table}
\begin{center}
\tiny \addtolength{\tabcolsep}{-0.5pt} \centering
\caption{Results of spectral graph partitioning.}
\begin{tabular}{|c|c|c|c|c|c|}
 \hline   Test Cases & $|V|$ & $\frac{|V_+|}{|V_-|}$ &  $T_{D}$ ($M_{D}$) &  $T_{I}$ ($M_{I}$) & Rel.Err.  \\
  \hline G3\_circuit & 1.6E6 &  1.35 & 52.3s (2.3G) & 7.6s (0.3G) & 2.2E-2  \\
 \hline thermal2 & 1.2E6 &  1.00 & 13.0s (0.9G) & 3.0s (0.2G) & 6.8E-4  \\
 \hline ecology2 & 1.0E6 &   1.03 & 12.1s (0.7G) & 3.4s (0.2G) & 8.9E-3  \\
 \hline tmt\_sym & 0.7E6 & 0.99& 10.2s (0.6G) & 1.9s (0.1G) &2.1E-2  \\
 \hline paraboli\_fem & 0.5E6 & 0.98 & 8.8s (0.4G) & 2.4s (0.1G) & 3.9E-2  \\
  \hline mesh\_1M & 1.0E6 & 1.01 & 10.2s (0.7G) & 1.7s (0.2G) & 3.3E-3  \\
  \hline mesh\_4M & 4.5E6 & 0.99 & 49.6s (3.0G) & 8.2s (0.7G) & 7.5E-3  \\
  \hline mesh\_9M & 9.0E6 & 0.99 & 138.5s (6.9G) & 13.3s (1.5G) & 7.8E-4  \\
  \hline
\end{tabular}\label{table:fiedler}
\end{center}
\end{table}
  \vspace{-0.cm}
\subsection{Sparsification of Other Complex networks}
\begin{table}
\begin{center}
\tiny \addtolength{\tabcolsep}{-0.5pt} \centering
\caption{Results of complex network sparsification.}
\begin{tabular}{|c|c|c|c|c|c|c|}
 \hline   Test Cases & $|V|$ & $|E|$  &  $T_{tot}$& $\frac{|E|}{|Es|}$ & $\frac{\lambda_1}{\tilde{\lambda}_1}$& ${T_{eig}^{o}}({T_{eig}^{s}})$  \\
%   \hline Reuters911 & 1.3E4 & 1.5E5 & 0.2s & 1.6E4 & $9\times$ & $21\times$& 2.1s (0.4s)\\
%   \hline pdb1HYS & 3.6E4 & 2.2E6 & 3.8s  & 7.1E4& $31\times$ & $8,981\times$& 10.5s (1.3s)\\
  \hline fe\_tooth & 7.8E4 & 4.5E5 & 3.0s  & $5\times$ & $8,050\times$& 14.5s (2.7s)\\
    \hline appu & 1.4E4 & 9.2E5 & 5.4s & $25\times$ & $13,624\times$&  2,400s (15s)\\
  \hline coAuthorsDBLP & 3.0E5 & 1.0E6 & 7.2s & $3\times$ & $1,364\times$& 2,047s (36s)\\
\hline auto & 4.5E5 & 3.3E6 &29.0s & $5\times$ & $48,190\times$&  N/A (54s)\\
  \hline RCV-80NN & 1.9E5 & 1.2E7 & 46.5s  & $36\times$ & $28,410\times$&  N/A (170s)\\
  \hline
\end{tabular}\label{table:uflsocial}
\end{center}
\end{table}
As shown in Table \ref{table:uflsocial},  a few finite element,   protein, data and social networks  have been  spectrally sparsified to achieve $\sigma^2 \approx 100$ using the proposed similarity-aware method. ``$T_{tot}$" is the total time for extracting the sparsifier, ``$\frac{\lambda_1}{\tilde{\lambda}_1}$" denotes the  ratio of  the largest generalized eigenvalues before and after adding off-tree edges into the spanning tree sparsifier, and  ${T_{eig}^{o}}({T_{eig}^{s}})$ denotes the time for computing the first ten eigenvectors of the original  (sparsified) graph Laplacians using the ``$eigs$" function in Matlab. Since spectral sparsifiers can well approximate the spectral (structural) properties of the original graph, the sparsified graphs can be leveraged for accelerating many numerical and graph-related tasks. For example, spectral clustering (partitioning) using the original ``RCV-80NN"  (80-nearest-neighbor) graph can not be performed on our server with $50 GB$ memory, while it only takes a few minutes  using the sparsified one.

\section{Conclusions}\label{conclusion}
For the first time, this paper introduces a similarity-aware spectral graph sparsification framework that can be immediately  leveraged to develop fast numerical and graph-related algorithms. Motivated by recent graph signal processing concepts, an iterative graph densification procedure based on  spectral embedding and filtering of off-tree edges has been proposed for extracting ultra-sparse yet spectrally-similar graph sparsifiers, which enables to flexibly trade off the complexity and spectral similarity of the sparsified graph  in numerical and graph-related applications. Extensive experiment results  have confirmed the   effectiveness and scalability of an iterative matrix solver and a spectral graph partitioning algorithm  for a variety of large-scale, real-world graph problems, such as VLSI and finite element analysis problems, as well as data  and social   networks.

\section{Acknowledgments}
This work is supported in part by  the National Science Foundation under Grants  CCF-1350206 and CCF-1618364.
%  \vspace{-0.03cm}
% \bibliographystyle{unsrt}
\bibliographystyle{abbrv}
{
\bibliography{dac18-spectralgraph}  % sigproc.bib is the name of the Bibliography in this case
%\scriptsize
}

\end{document}